\documentclass[universe,review,accept,moreauthors]{Definitions/mdpi} 
\firstpage{1} 
\pubvolume{11}
\issuenum{7}
\articlenumber{236}
\pubyear{2025}
\copyrightyear{2025}
\externaleditor{Gerald B. Cleaver} 
\datereceived{20 June 2025} 
\daterevised{11 July 2025} 
\dateaccepted{17 July 2025} 
\datepublished{19 July 2025} 
\hreflink{https://doi.org/\linebreak  10.3390/universe11070236} 

\Title{Evolving Wormholes in a Cosmological Background}

\TitleCitation{Evolving Wormholes in a Cosmological Background}

\Author{Mahdi 
 Kord Zangeneh $^{1,}$*\orcidA{} 
 and Francisco S. N. 
 Lobo $^{2,3}$\orcidB{}}
 \AuthorCitation{Kord 
 Zangeneh, M.; Lobo, F.S.N.}
\address{$^{1}$ \quad Physics Department, Faculty of Science, Shahid Chamran University of Ahvaz, Ahvaz 61357-43135, Iran\\
$^{2}$ \quad Instituto de Astrof\'{i}sica e Ci\^{e}ncias do Espa\c{c}o, Faculdade de Ci\^encias da 
 Universidade de Lisboa, Edif\'{i}cio C8, Campo Grande, 1749-016 Lisbon, Portugal; fslobo@fc.ul.pt\\
$^{3}$ \quad Departamento de F\'{i}sica, Faculdade de Ci\^{e}ncias, Universidade de Lisboa, Edifício C8, Campo Grande, 1749-016 Lisbon, Portugal}

\corres{Correspondence: mkzangeneh@scu.ac.ir}

\abstract{Wormholes are non-trivial topological structures that arise as exact solutions to Einstein's field equations, theoretically connecting distinct regions of spacetime via a throat-like geometry. While static traversable wormholes necessarily require exotic matter that violates the classical energy conditions, subsequent studies have sought to minimize such violations by introducing time-dependent geometries embedded within cosmological backgrounds. This review provides a comprehensive survey of evolving wormhole solutions, emphasizing their formulation within both general relativity and alternative theories of gravity. We explore key developments in the construction of non-static wormhole spacetimes, including those conformally related to static solutions, as well as dynamically evolving geometries influenced by scalar fields. Particular attention is given to the wormholes embedded into Friedmann–Lemaître–Robertson–Walker (FLRW) universes and de Sitter backgrounds, where the interplay between the cosmic expansion and wormhole dynamics is analyzed. We also examine the role of modified gravity theories, especially in hybrid metric--Palatini gravity, which enable the realization of traversable wormholes supported by effective stress--energy tensors that do not violate the null or weak energy conditions.
By systematically analyzing a wide range of time-dependent wormhole solutions, this review identifies the specific geometric and physical conditions under which wormholes can evolve consistently with null and weak energy conditions. These findings clarify how such configurations can be naturally integrated into cosmological models governed by general relativity or modified gravity, thereby contributing to a deeper theoretical understanding of localized spacetime structures in an expanding universe.}

\begin{document}




\section{Introduction}

Wormholes are hypothetical non-trivial topological structures that arise as exact solutions to Einstein's field equations, characterized by two distinct regions of spacetime connected through a throat-like geometry. These structures have been studied within the framework of general relativity and alternative theories of gravity due to their implications for the causal structure of spacetime and energy condition violations.
A foundational contribution to modern theoretical research on wormholes was made by Morris and Thorne~\cite{Morris:1988cz}, who formulated the concept of a static, spherically symmetric traversable wormhole. 
Their goal was to explore whether the Einstein field equations admitted geometries that permitted passage through the wormhole without encountering horizons or singularities. They demonstrated that keeping the wormhole throat open in a static configuration necessarily requires the violation of the \textit{null energy condition} (NEC) by the matter threading the wormhole.
More specifically, the~stress--energy tensor $T_{\mu\nu}$ associated with these wormhole geometries must satisfy $T_{\mu\nu} k^\mu k^\nu < 0$ for some null vector $k^\mu$, implying the presence of ``exotic matter''. This conclusion sparked extensive research into the nature and theoretical viability of such exotic sources, such as the fundamental role of energy conditions, which are important in the classical singularity theorems and the causal structure of general relativity~\cite{Hawking:1973uf,Wald:1984rg}, highlighting the significance of their violation in the context of~wormholes.

Subsequent investigations increasingly focused on reducing the dependence on exotic matter. In~particular, one~line of research has explored the possibility of dynamical or time-dependent spacetime geometries, wherein violations of energy conditions, such as the weak energy condition (WEC), may be restricted to localized regions or occur transiently. The~development of evolving or non-static wormhole solutions has therefore emerged as a significant and productive avenue of the theoretical research~\cite{Kar:1994tz, Kar:1995ss, Anchordoqui:1997du}. 
Thus, motivated both by the requirements of embedding wormholes into cosmological backgrounds and by the hypothesis that time evolution could improve or confine the extent of energy condition violations, researchers have constructed a variety of dynamical wormhole models. Among~the earliest contributions in this direction is the work of Kar~\cite{Kar:1994tz}, who investigated spacetimes that are conformally related to the static Morris–Thorne wormhole metric. In~this framework, the~introduction of a time-dependent conformal factor allows for the construction of non-static wormhole geometries that evolve over time while preserving the essential topological features of their static counterparts. It was shown that such evolving wormhole solutions can exist over finite, but~potentially arbitrarily large, intervals of cosmic time. Under~specific choices of the conformal factor and associated metric functions, the~resulting stress--energy tensor can be arranged to satisfy the WEC throughout the duration of the wormhole’s existence. This represents a significant theoretical development, as~static wormhole configurations invariably require stress--energy sources that violate at least the~WEC.

Expanding upon this framework, Kar and Sahdev~\cite{Kar:1995ss} developed a more general formalism for constructing spherically symmetric, time-dependent Lorentzian wormholes. Their analysis involved metrics characterized by time-dependent scale factors, allowing for the examination of evolving geometries under various cosmological expansion scenarios. The~authors explored several specific cases, including exponential and power-law scale factors, as~well as metrics incorporating additional compact extra dimensions. In~scenarios involving exponentially decaying compact dimensions, analogous to those encountered in Kaluza--Klein-type models, distinct dynamical behaviors of the wormhole throat were observed, highlighting how the extra-dimensional dynamics can significantly influence the wormhole's evolution and its compliance with energy conditions.
Anchordoqui~et~al.~\cite{Anchordoqui:1997du} extended the study of dynamic wormhole geometries by constructing solutions in the context of an expanding universe driven by scalar fields. Their analysis demonstrated that scalar field configurations with minimal coupling to gravity can support cosmologically evolving wormhole structures, wherein the spacetime exhibits non-trivial temporal behavior. These results highlight the viability of scalar fields as sources of stress--energy capable of sustaining transient or evolving wormholes in a cosmological setting without the need for exotic matter violating energy conditions at all~times.

The interplay between the local wormhole geometry and large-scale cosmological evolution was explored further by Roman~\cite{Roman:1992xj}, who examined Lorentzian wormholes embedded into a de Sitter spacetime. In~this framework, driven by a positive cosmological constant, Roman presented a novel classical metric describing a wormhole whose geometry inflates in the de Sitter background. The~study revealed that both the proper radial length and the throat radius of the wormhole experience exponential growth due to inflation. This mechanism suggests that early-universe inflation could potentially act as a natural enlargement process for Planck-scale wormholes, thereby rendering them macroscopic and possibly traversable at later times.
In a related study, Kim~\cite{Kim:1995xf} investigated wormhole solutions embedded within Friedmann–Lemaître–Robertson–Walker (FLRW) cosmologies, considering the full backreaction of the cosmological expansion on the wormhole geometry. The~analysis demonstrated that the global scale factor dynamics significantly affect the wormhole’s throat properties and its stability conditions. Depending on the behavior of the cosmic expansion, the~wormhole geometry may exhibit varying degrees of stability or transient behavior, suggesting non-trivial coupling between the local topological features and the universe's large-scale structure.
\textls[-15]{Much work has been conducted on time-dependent wormholes, and~rather than summarizing the literature here, we refer the reader to Refs.~\cite{Arellano:2006ex,Mirza:2006vr,Arellano:2008xw,Cataldo:2008pm,Cataldo:2008ku,Ebrahimi:2009ih,El-Nabulsi:2010sim,Farooq:2010rh,Cataldo:2011zn,Mehdizadeh:2012zz,Pan:2014oaa,Setare:2015iqa,Bhattacharya:2015oma,Bittencourt:2017yxq,Golchin:2019qch,Parsaei:2020hke,Mehdizadeh:2020nrw,KordZangeneh:2020jio,Bhattacharya:2021frx,Mehdizadeh:2021kgv,Heydarzade:2023ofq,Dutta:2024aum,Varsha:2025qgv}.}

Moreover, alternative theories of gravity, such as $f(R)$ gravity~\cite{Lobo:2009ip,Harko:2013yb,Pavlovic:2014gba,Godani:2018blx,Bronnikov:2010tt,Nashed:2019tuk,Bahamonde:2016ixz,DeBenedictis:2012qz,Godani:2019kgy,Godani:2020vqe,Mazharimousavi:2016npo,Sharif:2013dga}; 
hybrid metric--Palatini gravity~\cite{Capozziello:2012hr,Rosa:2018jwp}; teleparralel gravity~\cite{Boehmer:2012uyw,Mustafa:2021vqz}; linear curvature--matter couplings~\cite{MontelongoGarcia:2010xd,Garcia:2010xb}; $f(R,T)$ gravity, where $T$ is the trace of the energy momentum \linebreak  tensor~\cite{Azizi:2012yv,Zubair:2016cde,Moraes:2017mir,Rosa:2022osy,Yousaf:2017hjh,Moraes:2019pao,Moraes:2017rrv,Banerjee:2020uyi,Sahoo:2019aqz}; energy-momentum squared gravity~\cite{Moraes:2017dbs,Rosa:2023guo}; $f(R,L_m)$ gravity~\cite{Solanki:2023onp,Kavya:2023tjf}; non-metricity $f(Q)$ gravity~\cite{Banerjee:2021mqk,Calza:2022mwt}; $f({\cal T})$ gravity, where ${\cal T}$ is the torsion scalar~\cite{Sharif:2013lya,Jawad:2022aky}; higher dimensions~\cite{Mehdizadeh:2015dta,Zangeneh:2014noa}; Finsler--Randers geometry~\cite{Das:2023oli}; Einstein--Cartan gravity~\cite{Mehdizadeh:2017dhb}; modified Gauss--Bonnet gravity~\cite{ZeeshanGul:2024vem,Sharif:2014yea,Sharif:2015ffa}; and conformal Weyl gravity~\cite{Lobo:2008zu,Heydari-Fard:2023zfk}, amongst many other approaches to gravity, have provided frameworks where the effective stress--energy tensor includes geometric contributions that can mimic the exotic matter's behavior. These extensions allow for the realization of wormhole solutions without explicit violations of the classical energy conditions when interpreted in terms of effective fluid models.
A particularly promising framework is hybrid metric--Palatini gravity~\cite{Harko:2011nh,Harko:2018ayt}, which combines the features of both metric and Palatini formalisms. This theory naturally introduces extra scalar degrees of freedom that can mimic the effects of exotic matter~\cite{Capozziello:2012qt}. Several authors have explored wormhole~\cite{Bronnikov:2019ugl, Bronnikov:2020vgg}, black hole~\cite{Rosa:2020uoi, Chen:2020evr}, and~cosmological~\cite{Lima:2014aza, Lima:2015nma} solutions in this context. These studies demonstrate that hybrid gravity can support traversable wormholes without explicit energy condition~violations.

Hybrid metric--Palatini gravity exhibits a broad range of astrophysical applications. In~particular, potential observational signatures of this theory at the Galactic Center have been investigated. In~Ref.~\cite{Borka:2015vqa}, the~authors analyzed the orbital precession of the S2 star around the supermassive compact object at the Milky Way's center, finding excellent agreement between simulated orbits in hybrid gravity and astronomical observations. This work was extended in Ref.~\cite{Borka:2022sot}, where the parameters of the hybrid metric--Palatini model were estimated using the Schwarzschild precession of S2, S38, and~S55. Various values for the bulk mass density of extended matter were considered, with the~results again being consistent with observational data.~Ref.~\cite{BorkaJovanovic:2021uew} further constrained hybrid gravity using the fundamental plane of elliptical galaxies in the weak-field regime. This~study showed that hybrid gravity can reproduce the velocity dispersion and structural properties of elliptical galaxies without invoking dark matter.
Additionally, alternative modifications to General GR have been proposed to explain specific astrophysical phenomena. For~instance, a~Yukawa-type modification of the Newtonian potential in the weak-field limit was investigated in Ref.~\cite{Jovanovic:2022twh}. Using the Parametrized Post-Newtonian (PPN) formalism, the~authors simulated S-star orbits at the Galactic Center under both GR and Yukawa gravity, concluding that the latter offers a viable framework for constraining modified gravity parameters and testing alternative gravitational~theories.

In Ref.~\cite{KordZangeneh:2020ixt}, the~authors investigated the dynamics of time-dependent traversable wormhole configurations embedded within a Friedmann–Lemaître–Robertson–Walker (FLRW) cosmological background, in~the framework of the scalar--tensor representation of hybrid metric--Palatini gravity. Within~this theoretical context, the~authors derived the energy-momentum tensor components corresponding to the matter threading the wormhole geometry, expressing them in terms of the scalar field, the~cosmological scale factor, the~wormhole shape function, and~other background cosmological quantities.
To obtain explicit solutions, they considered a barotropic equation of the state for the background matter, characterized by a linear relationship between pressure and energy density. Under~this assumption, this~study demonstrates the existence of specific classes of wormhole solutions for which the matter content satisfies both the null energy condition (NEC) and the weak energy condition (WEC) at all times during the cosmological~evolution.

In this work, we investigate evolving wormholes within cosmological backgrounds by considering both general relativity and extended gravity theories, with~a particular focus on hybrid metric--Palatini gravity. Our aim is to determine the conditions under which wormholes can remain traversable and physically viable across cosmic epochs. By~embedding dynamical wormhole metrics into FLRW backgrounds and analyzing their evolution under realistic matter sources and modified gravitational couplings, we explore new pathways for integrating localized topological structures into a consistent cosmological framework. Remarkably, we will see that several resulting spacetime geometries satisfy the NEC and WEC throughout the entire temporal evolution. These results represent important progress in identifying physically viable wormhole models that are compatible with generalized theories of gravity and are supported by non-exotic matter~sources.

This paper is organised in the following manner: In Section~\ref{sectionII}, we present the spacetime metric and the gravitational field equations and~analyze several specific solutions, including an evolving wormhole in a flat FRW universe. In~Section~\ref{sectionIII}, we present a wormhole in a time-dependent inflationary background and~analyze the properties of the solutions and the NEC violation. In~Section~\ref{sectionIV}, we analyze evolving cosmological wormholes in hybrid metric--Palatini gravity. Finally, in~Section~\ref{section:conclusion}, we summarize and discuss our resutls and~conclude.

\section{Evolving Wormholes in a Cosmological~Background}
\label{sectionII}
\unskip

\subsection{The Spacetime Metric and the Stress--Energy~Tensor}


Consider the metric element of a wormhole in a cosmological background given
by
\begin{equation}  \label{evolvingWHmetric}
ds^{2} = \Omega ^{2}(t) \left[- e ^{2\Phi(r)}\, dt^{2} + {\frac{{dr^{2}}}{{%
1-kr^2- \frac{b(r)}{r}}}} + r^2 \,\left(d\theta ^2+\sin ^2{\theta} \, d\phi
^2 \right) \right]
\end{equation}
where $\Omega ^{2}(t)$ is the conformal factor, which is finite and positive
definite throughout the domain of $t$. It is also possible to write the
metric (\ref{evolvingWHmetric}) using ``physical time'' instead of
``conformal time'', by~replacing $t$ with $\tau = \int \Omega (t)dt$ and
therefore $\Omega (t)$ with $R(\tau)$, where the latter is the functional form
of the metric in the $\tau$ coordinate~\cite{Kar:1994tz,Kar:1995ss}. We
shall use `conformal time' in the present analysis, reserving $\tau$ for the
specific cosmological models considered further ahead. When the form
function and the redshift function vanish, $b(r)\rightarrow 0$ and $%
\Phi(r)\rightarrow 0$, respectively, the~metric (\ref{evolvingWHmetric})
becomes the FRW metric. As~$\Omega(t)\rightarrow \mathrm{const}$ and $%
k\rightarrow 0$, it approaches the static wormhole~metric.


To analyze the stress--energy tensor of the wormhole described by Equation (\ref%
{evolvingWHmetric}), consider the orthonormal basis vectors defined by 
\begin{eqnarray}
\mathbf{e}_{\hat{t}}&=&\Omega ^{-1}\,e^{-\Phi}\,\mathbf{e}_{t} ,  \notag \\
\mathbf{e}_{\hat{r}}&=&\Omega ^{-1}\,(1-b/r)^{1/2}\,\mathbf{e}_{r},  \notag
\\
\mathbf{e}_{\hat{\theta}}&=&\Omega ^{-1}\,r^{-1}\,\mathbf{e}_{\theta}, 
\notag \\
\mathbf{e}_{\hat{\phi}}&=&\Omega ^{-1}\,(r\,\sin\theta)^{-1}\,\mathbf{e}%
_{\phi}.
\end{eqnarray}

This 
 basis represents the proper reference frame of a set of observers who
always remain at rest at constant $r,\,\theta,\,\phi$. The~Einstein field
equation will be written as\linebreak   $G_{\hat{\mu}\hat{\nu}}=R_{\hat{\mu}\hat{\nu}} -{%
\frac{1}{2}}g_{\hat{\mu}\hat{\nu}}R ={8\pi}T_{\hat{\mu}\hat{\nu}}\,,$ so
that any cosmological constant terms will be incorporated as part of the
stress--energy tensor $T_{\hat{\mu}\hat{\nu}}$. The~components of the
stress--energy tensor $T_{\hat{\mu}\hat{\nu}}$ are given by
\begin{equation}
T_{\hat{t}\hat{t}}=\rho(r,t)\,, \qquad T_{\hat{r}\hat{r}}=-\tau(r,t) \,,
\qquad T_{\hat{t}\hat{r}}=-f(r,t) \,, \qquad T_{\hat{\phi}\hat{\phi}}=T_{%
\hat{\theta}\hat{\theta}}=p(r,t) \,,
\end{equation}
with
\begin{eqnarray}
\rho(r,t)&=&\frac{1}{8\pi}\,\frac{1}{\Omega^2}\,\left[ 3 e^{-2\Phi}\,\left(%
\frac{\dot {\Omega}}{\Omega}\right)^2 +\left(3k+\frac{b ^{\prime }}{r^2}
\right) \right] \,,  \label{dynTtt} \\
\tau(r,t)&=&-\frac{1}{8\pi}\,\frac{1}{\Omega ^{2}}\;\left\{ e^{-2\Phi(r)}\,%
\left[{\left ({\frac{{\dot {\Omega}} }{{\Omega}}}\right )^{2}}-2\,\frac{%
\ddot {\Omega}}{\Omega} \right] - \left[k+ {\frac{b }{{r^{3}}}}-2\,\frac{%
\Phi^{\prime }}{r}\,\left(1-kr^2-\frac{b}{r}\right) \right] \right\} \,, \\
f(r,t)&=&-\frac{1}{8\pi}\, \left[2\,\frac{\dot{\Omega}}{\Omega^3}%
\;e^{-\Phi}\,\Phi^{\prime }\left(1-kr^2-\frac{b}{r} \right)^{1/2} \right] \,,
\\
p(r,t)&=&\frac{1}{8\pi}\, \frac{1}{\Omega ^{2}}\;\Bigg\{ e^{-2\Phi(r)}\,%
\left[{\left ({\frac{{\dot {\Omega}} }{{\Omega}}}\right )^{2}}-2\,\frac{%
\ddot {\Omega}}{\Omega} \right] + \left(1-kr^2-\frac{b}{r}\right) \times 
\notag \\
&&\hspace{1cm}\times\left[\Phi ^{\prime \prime }+ (\Phi^{\prime 2 }- \frac{%
2kr^3+b^{\prime }r-b}{2r(r-kr^3-b)}\Phi^{\prime }- \frac{2kr^3+b^{\prime }r-b%
}{2r^2(r-kr^3-b)}+\frac{\Phi^{\prime }}{r} \right] \Bigg\} \,.  \label{dyn-p}
\end{eqnarray}
The overdot denotes a derivative with respect to $t$ and~the prime a
derivative with respect to $r$. The~physical interpretations of $%
\rho(r,t),\,\tau(r,t)\,,f(r,t)$, and~$p(r,t)$ are the following: the energy
density, the~radial tension per unit area, the energy flux in the (outward)
radial direction, and~lateral pressures as measured by observers stationed
at constant $r,\,\theta,\,\phi$, respectively. The~stress--energy tensor has
a non-diagonal component due to the time dependence of $\Omega(t)$ and/or
the dependence of the redshift function on the radial coordinate. The~stress--energy tensor of an imperfect fluid was analyzed in \cite%
{Anchordoqui:1997du}.

To analyze the evolving wormhole, one chooses $\Phi(r)$ and $b(r)$ to give a
reasonable wormhole at $t=0$, which is assumed to be the onset of the
evolution. The~radial proper length through the wormhole between any two
points $A$ and $B$ at any $t=\mathrm{const}$ is given by
\begin{equation}
l(t)=\pm \Omega(t) \int_{r_A}^{r_B} {\frac{dr}{{(1-kr^2-b/r)^{1/2}}}}\,,
\label{evolvproperdistance}
\end{equation}
which is just $\Omega(t)$ times the initial radial proper~separation.

To see that the ``wormhole'' form of the metric is preserved with time,
consider an embedding of a $t=\mathrm{const}$ and $\theta=\pi/2$ slice (without a significant loss of generality) of
the spacetime given by Equation~(\ref{evolvingWHmetric}) into~a flat three-dimensional
Euclidean space with the~metric
\begin{equation}
ds^2=d{\bar{z}}^2+d{\bar{r}}^2+{\bar{r}}^2\,{d\phi}^2\,.  \label{barredslice}
\end{equation}
The metric of the wormhole slice is
\begin{equation}  \label{slice}
ds^2={\frac{\Omega^2(t)\,{dr^2}}{{(1-kr^2-b(r)/r)}}} + \Omega^2(t)\,
r^2\,d\phi^2\,.
\end{equation}
Comparing the coefficients of ${d\phi}^2$, one has
\begin{eqnarray}
\bar{r}&=&{\Omega(t)\,r}\big|_{t=\mathrm{const}} \,,  \label{coef1:phi} \\
{d\bar{r}}^2&=&\Omega(t)^2\,{dr}^2\big|_{t=\mathrm{const}} \,.
\label{coef2:phi}
\end{eqnarray}
Note that it is important to keep in mind, in~particular when considering
derivatives, that Equations~(\ref{coef1:phi}) and (\ref{coef2:phi}) do not represent
a ``coordinate transformation'' but~rather a ``rescaling'' of the $r$
coordinate on each $t=\mathrm{constant}$ slice.

With respect to the ${\bar{z}},{\bar{r}},\phi$ coordinates, the~``wormhole''
form of the metric will be preserved if the metric on the embedded slice has
the form
\begin{equation}  \label{WHslice}
ds^2={\frac{{d{\bar{r}}^2}}{{[1-{\bar{b}(\bar{r})/{\bar{r}}]}}}} + {\bar{r}}%
^2{d\phi}^2\,,
\end{equation}
where $\bar{b}(\bar{r})$ has a minimum at some $\bar{b}(\bar{r}_0)=\bar{r}_0$%
. Equation~(\ref{slice}) can be rewritten in the form of Equation~(\ref{WHslice})
by using Equations~(\ref{coef1:phi}) and (\ref{coef2:phi}) and
\begin{equation}
\bar{b}(\bar{r})=\Omega(t)\,[kr^3+b(r)].  \label{bar:b}
\end{equation}
The evolving wormhole will have the same overall size and shape relative to
the ${\bar{z}},{\bar{r}},\phi$ coordinate system as~the initial wormhole
had relative to the initial $z,r,\phi$ embedding space coordinate system.
This due to the fact that the embedding space corresponds to $z,r$
coordinates which ``scale'' with time (each embedding space corresponds to a
particular value of $t=\mathrm{constant}$). Following the embedding
procedure outlined in~\cite{Morris:1988cz}, using Equations~(\ref{barredslice})~and~(\ref{WHslice}), one deduces that
\begin{equation}
{\frac{{d{\bar{z}}}}{{d{\bar{r}}}}} =\pm\left({\frac{{\bar{r}}}{{\bar{b}(%
\bar{r})}}}-1\right)^{-1/2} ={\frac{{dz}}{{dr}}} \,.  \label{barredembedding}
\end{equation}
Equation~(\ref{barredembedding}) implies
\begin{equation}
\bar{z}(\bar{r})=\pm\int{\frac{{d\bar{r}}}{{(\bar{r}/{\bar{b}(\bar {r})}%
-1)^{1/2}}}}= \pm \Omega(t)\,\int{\left(\frac{r-kr^3-b}{b+kr^3}\right)^{-1/2}%
} \,dr =\pm \Omega(t)\,z(r)\,.  \label{embed:relation}
\end{equation}
Therefore, we see that the relation between the embedding space at any time $%
t$ and the initial embedding space at $t=0$ is, from~Equations~(\ref{coef2:phi})
and (\ref{embed:relation}),
\begin{equation}
ds^2=d{\bar{z}}^2+d{\bar{r}}^2+{\bar{r}}^2\,{d\phi}^2
=\Omega^2(t)\,[dz^2+dr^2+r^2{d\phi}^2]\,.
\end{equation}
Relative to the ${\bar{z}},{\bar{r}},\phi$ coordinate system, the wormhole
will always remain the same size, as~the scaling of the embedding space
compensates for the evolution of the wormhole. However, the~wormhole will
change size relative to the initial $t=0$ embedding~space.

Writing the analog of the ``flaring-out condition''
for the evolving
wormhole, we have
\begin{equation}
{\frac{{d\,^2{\bar{r}(\bar{z})}}}{{d{\bar{z}}^2}}}>0\,,
\end{equation}
at or near the throat. From~Equations~(\ref{coef1:phi}), (\ref{coef2:phi}), (\ref%
{bar:b}), and~(\ref{barredembedding}), it follows that
\begin{equation}
{\frac{{d\,^2{\bar{r}(\bar{z})}}}{{d{\bar{z}}^2}}} =\frac{1}{\Omega(t)}\,{%
\frac{{b-b^{\prime 3}}}{{2(b+kr^3)^2}}} =\frac{1}{\Omega(t)}\,{\frac{{%
d\,^2r(z)}}{{dz^2}}}>0\,,  \label{barred:flareout}
\end{equation}
at or near the throat. Taking into account Equations~(\ref{coef1:phi}) and (\ref%
{bar:b}), and~\begin{equation}
{\bar{b}}^{\prime }(\bar{r})={\frac{{d\bar{b}}}{{d\bar{r}}}} =b^{\prime }(r)=%
{\frac{{db}}{{dr}}}\,,
\end{equation}
one may rewrite the right-hand side of Equation~(\ref{barred:flareout}) relative
to the barred coordinates as
\begin{equation}
{\frac{{d\,^2{\bar{r}(\bar{z})}}}{{d{\bar{z}}^2}}} =\left({\frac{{\bar{b}-{%
\bar{b}}^{\prime }\bar{r}}}{{2{\bar{b}}^2}}}\right)>0\,,
\label{barred:flareout2}
\end{equation}
at or near the throat. One verifies that using the barred coordinates, the~flaring-out condition, Equation~(\ref{barred:flareout2}), has the same form as that for
the static~wormhole.

\subsubsection{$\Phi=0$ Wormholes}

In particular, the~analysis is simplified considerably considering a null
redshift function, $\Phi(r)=0$. The~stress--energy tensor components reduce
to
\begin{eqnarray}
\rho(r,t)&=&\frac{1}{8\pi}\,\frac{1}{\Omega^2}\,\left[ 3\,\left(\frac{\dot {%
\Omega}}{\Omega}\right)^2 +3k+\frac{b^{\prime }}{r^2} \right] \,,
\label{FRW:WHrho} \\
\tau(r,t)&=&\frac{1}{8\pi}\,\frac{1}{\Omega ^{2}}\;\left[ {-\left ({\frac{{%
\dot {\Omega}} }{{\Omega}}}\right )^{2}}+2\,\frac{\ddot {\Omega}}{\Omega} +
k+ {\frac{b }{{r^{3}}}} \right] \,,  \label{FRW:WHtau} \\
f(r,t)&=&0 \,, \\
p(r,t)&=&\frac{1}{8\pi}\, \frac{1}{\Omega ^{2}}\;\left[ {\left ({\frac{{%
\dot {\Omega}} }{{\Omega}}}\right )^{2}}-2\,\frac{\ddot {\Omega}}{\Omega} - 
\frac{2kr^3+b^{\prime }r-b}{2r^3} \right] \,.  \label{FRW:WHp}
\end{eqnarray}

Considering the energy conditions, the~WEC $(T_{\mu \nu } U^{\mu } U^{\nu }
\ge 0$, $\forall$ timelike $U^{\mu })$ reduces to the following inequalities
for the case of a diagonal energy-momentum tensor \cite%
{Kar:1994tz,Kar:1995ss}:
\begin{equation}
\rho \ge 0,\quad \rho -\tau \ge 0 ,\quad \rho + p \ge 0 \quad \forall (r,t)
\,.
\end{equation}
Taking into account Equations~(\ref{FRW:WHrho}), (\ref{FRW:WHtau}), and (\ref%
{FRW:WHp}), one can write down three inequalities which have to be satisfied
if the WEC is not to be violated, given by
\begin{eqnarray}
\frac{1}{\Omega ^2}\left [3{{\left ({\frac{{\dot {\Omega}} }{{\Omega}}}%
\right )^{2}}}+3k+{\frac{{b ^{\prime}r} }{{r^3}}}\right ] &\ge &0 \,,
\label{WECrho} \\
{\frac{2}{{\Omega ^{2}}}}\left [-{\frac{{\ddot {\Omega}} }{{\Omega}}} + 2{%
\left ({\frac{{\dot {\Omega}} }{{\Omega}}}\right )^{2}} +k- {\frac{{b-{%
b^{\prime}} r}}{{2r^{3}}}} \right ] &\ge &0 \,,  \label{WECrho+tau} \\
{\frac{2}{{\Omega ^{2}}}}\left [-{\frac{{\ddot {\Omega}} }{{\Omega}}} + 2{%
\left ({\frac{{\dot {\Omega}} }{{\Omega}}}\right )^{2}}\right ] +k+{\frac{{b+%
{b^{\prime}} r} }{{2r^{3}\Omega^{2}}}} &\ge &0 \,.  \label{WECrho+p}
\end{eqnarray}

However, we are essentially interested in the flat universe analysis,
with $k=0$, in~which Equations~(\ref{WECrho})--(\ref{WECrho+p}) reduce to
\begin{eqnarray}
{\frac{2}{{\Omega ^{2}}}}\left [{\frac{3}{2}}{{\left ({\frac{{\dot {\Omega}} 
}{{\Omega}}}\right )^{2}}}+{\frac{{b ^{\prime}r} }{{2r^{3}}}}\right ] &\ge
&0 \,,  \label{flatWECrho} \\
{\frac{2}{{\Omega ^{2}}}}\left [-{\frac{{\ddot {\Omega}} }{{\Omega}}} + 2{%
\left ({\frac{{\dot {\Omega}} }{{\Omega}}}\right )^{2}} - {\frac{{b-{%
b^{\prime}} r}}{{2r^{3}}}} \right ] &\ge &0 \,,  \label{flatWECrho+tau} \\
{\frac{2}{{\Omega ^{2}}}}\left [-{\frac{{\ddot {\Omega}} }{{\Omega}}} + 2{%
\left ({\frac{{\dot {\Omega}} }{{\Omega}}}\right )^{2}}\right ] +{\frac{{b+{%
b^{\prime}} r} }{{2r^{3}\Omega^{2}}}} &\ge &0 \,.  \label{flatWECrho+p}
\end{eqnarray}

Note that the inequality (\ref{flatWECrho}) is satisfied if $b^\prime \ge 0$
irrespective of the geometry being static or dynamic. However, if~it is time-dependent, then the inequality (\ref{flatWECrho}) is satisfied even for a
case where $b^{\prime} \le 0$, where one obtains the restriction
\begin{equation}
{\frac{{\vert {b^{\prime}} \vert}}{r^{2}}} \le 3{\left({\frac{\dot \Omega }{%
\Omega}} \right)}^{2} \,.  \label{flatWECrho:restriction}
\end{equation}
For every $t=\mathrm{constant}$ slice, the inequality (\ref%
{flatWECrho:restriction}) has to hold, which means
\begin{equation}
{\frac{{\vert {b^{\prime}} \vert}}{r^{2}}} \le \mathrm{min} \left [ 3{\left({%
\frac{\dot \Omega }{\Omega}}\right)} ^{2}\right ] \,,
\end{equation}
where $\mathrm{min}$ denotes the minimum value of the function in the given
time~interval.

The inequality (\ref{flatWECrho+tau}) can never be satisfied for the static
Morris--Thorne wormhole~\cite{Morris:1988cz}. However, for~the present
evolving geometry with $b^\prime \ge 0$, it can be obeyed by~imposing
\begin{equation}
F(t) \ge {\frac{{b-{b^{\prime}} r}}{{2r^{3}}}} \,,  \label{dyn-WEC}
\end{equation}
where for~simplicity, the~definition
\begin{equation}
F(t)=\left [2\left( \frac{\dot{\Omega}}{\Omega} \right)^2 -\frac{\ddot{\Omega%
}}{\Omega}\right ]
\end{equation}
is used. The~inequality (\ref{dyn-WEC}) implies that the value of $%
(b-b^\prime r)/r^{3}$ for all $r$ must be less than or equal to the minimum
value of the function $F(t)$, in~the domain of $t$. Note that the
imposition $F(t)>0$ is also necessary, as~$(b-b^\prime r)/r^{3}>0$, due to
the flaring-out condition. Therefore, one verifies from this analysis that
dynamic spherically symmetric wormhole geometries can exist without violation of the WEC. However, these WEC-satisfying evolving wormholes may only exist
for finite intervals of time due to the fact that $\Omega(t)$ is finite
everywhere and satisfies the condition (\ref{dyn-WEC}).

One can also use the Raychaudhuri equation to verify the non-violation of
the WEC for dynamic wormholes~\cite{Kar:1995ss}. Recall that the
Raychaudhuri equation for a congruence of null rays {\cite{Wald:1984rg} is
given by
\begin{equation}
{\frac{d\hat{\theta}}{d\lambda}}=-\frac{1}{2}{\hat{\theta}}^{2} -R_{\mu\nu}\,%
{k}^{\mu}\, {k}^{\nu} -2{\hat{\sigma}}^{2} +2\hat{\omega}^2 \,,
\label{nullRaychaud}
\end{equation}
where $\hat{\theta}$ is the expansion of the congruence of null rays; $\hat{%
\sigma}$ and $\hat{\omega}$ are the shear force~
and vorticity of the geodesic
bundle, respectively, which are zero for this case. From~the Einstein field
equation, we have $R_{\mu\nu}\, k^{\mu}\,k^{\nu}=T_{\mu\nu}\,k^{\mu}\,k^{\nu}$
for all null $k^{\mu}$. So, if $T_{\mu\nu}\,k^{\mu}\,k^{\nu} \ge 0$, we have $%
\hat{\theta}^{-1}\ge {\hat{\theta}}_{0}^{-1}+\frac{\lambda}{2}$ according to (\ref%
{nullRaychaud}). Thus, if $\hat{\theta}$ is negative anywhere, it has to go to 
$-\infty$ at a finite value of the affine parameter, $\lambda$; i.e.,~the
bundle must necessarily come to a focus. 
Note that the physical meaning of $\hat{\theta}$ being negative is simply that the volume of a small bundle of neighbouring geodesics is decreasing, namely that the geodesics are converging or focusing, which implies that gravity is attractive.
In the case of a static wormhole,
the expansion $\hat{\theta}$ is given by
\begin{equation}
\hat{\theta} = \frac{2\beta}{r} \frac{dr}{dl} \,,
\end{equation}
where $\beta$ is a positive quantity. For~$l<0$, $dr/dl$ is negative, so then
is $\hat{\theta}$ negative. However, $\hat{\theta} \rightarrow -\infty$ only
if $r \rightarrow 0$ since $dr/dl$ is always finite. Therefore, either the
wormhole has a vanishing throat radius, which renders it non-traversable,
or the WEC is violated. }

For the evolving case, the~expansion is given by
\begin{equation}
\hat{\theta} = \frac{2\beta}{R(\tau)}\left (\frac{dR(\tau)}{d\tau}+\frac{1}{r%
}\,\frac{dr}{dl} \right) \,.
\end{equation}
Real time $\tau$ has now been used, and~as long as $dR/d\tau >(1/r)|dr/dl|$,
i.e., the~wormhole is opening out fast enough, $\hat{\theta}$ is never
negative. Thus, the~fact that the bundle does not focus no longer implies
that the WEC is~satisfied.

\subsubsection{Specific~Examples}

For simplicity, in~the examples that follow, the~specific choice of the form
function given by $b(r) =r_0$ is used (other choices of $b(r)$ may also be
used) \cite{Kar:1994tz,Kar:1995ss}. Thus, the~WEC, condition (\ref{dyn-WEC}%
), imposes the following restriction on $r_{0}$:
\begin{equation}
r_{0}^{2} \ge \mathrm{max} {\left ({\frac{1}{{F(t)}}}\right )} \,,
\end{equation}
which restricts the minimum possible value of the throat radius of the
wormhole.

\begin{enumerate}

\item $\Omega (t)=(C - \omega t)^{-1}$.

Note that this is essentially the case of an inflationary wormhole universe, as
discussed by Roman~\cite{Roman:1992xj}, which shall briefly be discussed
further ahead. The~fundamental idea was to obtain a macroscopic model of a
wormhole which emerged from inflation. This choice of the scale factor leads
to $F(t) = 0$, implying that the WEC is violated for all~times.

\item $\Omega (t) = \sin {\omega t} $.

This is essentially the scale factor used in closed FRW cosmological models.
The expression for $F(t)$ is given by
\begin{equation}
F(t) = 2{\omega}^{2}( 2\cot ^{2}{\omega t} + 1) \,.
\end{equation}
$F(t)$ has a minimum at $\omega t = \pi / 2 $; thus, the constraint on the
allowed values of $r_0$ is $r_0^2 {\omega}^2 \ge 1/2$. The~lifetime of this
wormhole universe is $\pi / \omega$, and~the time interval for which this
universe can exist without collapsing into a singularity is $m\pi /\omega< t
< (m+1)\pi / \omega $.

\item $\Omega (t) = {(\omega t)}^{\nu}$, $\nu$ integral or~fractional.

This is an important case, as~the scale factors that arise in
dust-filled or radiation-dominated FRW cosmologies with flat spacelike
sections are obtained by considering specific values of $\nu$. For~a
generic $\nu $, the expression for $F(t)$ is given as
\begin{equation}
F(t) = {\frac{{2\nu (\nu +1)}}{{t^{2}}}} \,,
\end{equation}
and $F(t) \rightarrow 0$ as $t \rightarrow \pm \infty $. The~constraint on $%
r_{0}$ is dependent on $t$, i.e,
\begin{equation}
r_0^{2} \ge \mathrm{max}{\left ({\frac{t^{2}}{{2\nu(\nu +1)}}}\right )} \,.
\end{equation}
This evolving wormhole can exist only for a finite interval of time. The~lower bound on $r_{0}$ is decided by the maximum time $t$ up to which we
wish the wormhole to exist with the matter threading the geometry satisfying
the WEC.
\end{enumerate}


Consider now a special class of scale factors which exhibit ``flashes'' of
WEC violation, where the matter threading the wormhole violates the energy
conditions for small intervals of time. It is interesting to note that one
can consider specific cases in~which the intervals of WEC violation can be
chosen to be very small. Consider the following examples:

\begin{enumerate}

\item $\Omega (t) = \sin {\omega t} + a\,,\;$ $a>1$.

This scale factor is reminiscent of the `bounce'-type solutions in
cosmology which were constructed in order to avoid the Big Bang
singularity. Thus, the~function $F(t)$ is given by
\begin{equation}
F(t) = 2{\omega}^{2}\left [{\frac{{2-{\sin}^{2}{\omega t} + a\sin {\omega t}}%
}{{({\sin \omega t +a})^{2}}}}\right ] \,,
\end{equation}
which is plotted in Figure~\ref{fig:evolvWH1}~\cite{Kar:1995ss}. $F(t)$
takes negative values periodically, for~the values of $\omega t$ given by $%
(2n+1)\pi /2$. This implies the violation of the WEC in the respective
interval. Considering small values of the parameter $a$, one verifies that the
interval for which $F(t)$ is negative becomes significantly~shortened.~

\begin{figure}[H]
\includegraphics[width=2.6in]{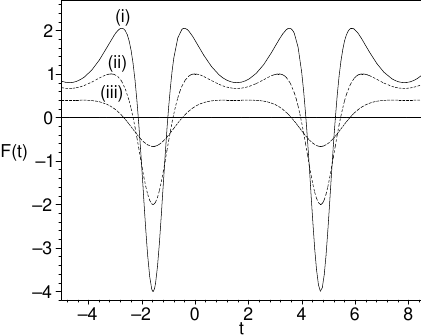}
\caption[Evolving wormhole exhibiting flashes of WEC violation: I]{A plot
of $F(t)$ vs. $t$ for $\Omega(t)=\sin(\protect\omega\,t)+a$. Only the case of 
$\protect\omega=1$ is considered, with~ respective values of (i) $a=1.5$%
, (ii) $a=2$, and (iii) $a=4$.}
\label{fig:evolvWH1}
\end{figure}

\item $\Omega (t)=(t^2+a^2)/(t^2+b^2)\,,\;$ $b>a>0$.

This choice of the scale factor implies that asymptotically, the geometry
becomes a static wormhole. The~function $F(t)$ is given by
\begin{equation}
F(t) = {\frac{{4(b^{2}-a^{2})(3t^{2}-a^{2})}}{{(t^{2}+b^{2})({t^{2}+a^{2}}%
)^{2}}}} \,.
\end{equation}
which is plotted in Figure~\ref{fig:evolvWH2}~\cite{Kar:1995ss}. At~$t=\pm
a/\sqrt 3$, we have $F(t) = 0$. For~$-a/\sqrt 3 < t < a/\sqrt 3$, $F(t)$ is
negative,~
implying the violation of the WEC.
\end{enumerate}


\begin{figure}[H]
\includegraphics[width=2.6in]{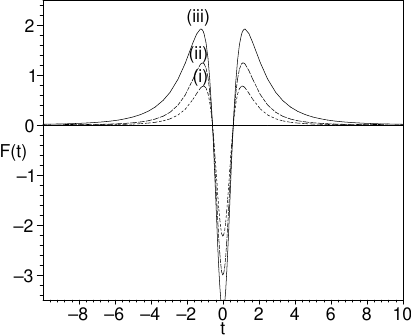}
\caption[Evolving wormhole exhibiting flashes of WEC violation: II]{A plot
of $F(t)$ vs. $t$ for $\Omega(t)=(t^2+a^2)/(t^2+b^2)$, $b>a>0$, with~$a=1$.
Cases of (i) $b=3$, (ii) $b=2$, and (iii) $b=1.5$ are considered,
respectively.}
\label{fig:evolvWH2}
\end{figure}
\unskip

\subsection{A Wormhole in a Flat FRW~Universe}

Consider now a wormhole evolution in a flat $k=0$ FRW universe~\cite%
{Kar:1994tz,Kar:1995ss}. One chooses the scale factor to be identical to
that in either the matter- or radiation-dominated FRW model \cite%
{Kim:1995xf}. Thus, the~metric is given by
\begin{equation}
	ds^{2} = -d{\tau}^{2} + \tau^n\left [ \frac{dr^2}{1-b(r)/r} + +r^2
	\,(d\theta ^2+\sin ^2{\theta} \, d\phi ^2 )\right ] \,.  \label{evolvWH-FRW}
\end{equation}
where real time is now used. $n$ takes the values $1/2$ and $2/3 $ for the
radiation- and matter-dominated cases of the flat FRW universes,
respectively. The~stress--energy scenario for the metric (\ref{evolvWH-FRW})
is the following:
\begin{eqnarray}
	\rho (r,\tau) &=& {\frac{b^{\prime }}{{{\tau}^{2n}r^{2}}}} + {\frac{{3n^{2}}%
		}{{\tau}^2}} \,, \\
	p_{1}({r,\tau}) &=& -{\frac{b}{{{\tau^{2n}}r^3}}} - {\frac{{n(3n-2)}}{{{\tau}%
				^2}}} \,, \\
	p_{2}(r,\tau) &=& {\frac{{b-b^{\prime }r}}{{{2r^3}{\tau^{2n}}}}} - {\frac{{%
				n(3n-2)}}{{\tau^2}}} \,.
\end{eqnarray}
For $r \rightarrow \infty $, one obtains $\rho = {{\rho}_{FRW}} =
4/(3\tau^2) $, $p_1= p_2=0$ for dust and $\rho/3 = p_1= p_2=1/(4\tau^2)$ for
pure~radiation.

The WEC inequality $\rho + p_{1} \ge 0 $ reduces to the following:
\begin{equation}
	{\frac{{b^{\prime}r - b}}{{{2r^{3}}{({\tau})^{2n}}}}} + {\frac{{2n}}{{({c\tau%
				})^{2}}}} \ge 0 \,.
\end{equation}
One may now consider several cases for the form function $b(r)$ and~analyze
the conditions that sprout therefrom. However, we shall not attempt this
endeavor here. The~particular case for $b(r)=r_0$ is analyzed in \cite%
{Kar:1994tz,Kar:1995ss}.

\section{Wormholes in an Inflationary~Background}\label{sectionIII}
\unskip

\subsection{The Metric and the Stress--Energy~Tensor}

A particularly interesting case of metric (\ref{evolvingWHmetric}) is
that of a wormhole in a time-dependent inflationary background considered
by Thomas Roman~\cite{Roman:1992xj}. We shall consider real time $\tau$
defined by $\tau=\int\,\Omega(t)\,dt$ and~replace $\Omega(t)$ with $R(\tau)$.
However, for~notational convenience, we shall use the $t$ coordinate (which
below is interpreted as real time). Thus, the~metric is given by
\begin{equation}
ds^2=-e^{2\Phi(r)}dt^2+e^{2\chi t}\left[{\frac{{dr^2}}{{(1-b(r)/r)}}} +r^2({%
d\theta}^2+ sin^2\theta\,{d\phi}^2)\right]\,,  \label{infWHmetric}
\end{equation}
where the spatial part of the metric is multiplied by a de Sitter scale
factor $R(t)=e^{2\chi t}$, with~$\chi=\sqrt{\Lambda/3}$; $\Lambda$ is the
cosmological constant~\cite{Heydarzade:2023ofq}. The~coordinates $r,\theta,\phi$ have the same
geometrical interpretation as that in the Morris--Thorne case. For~$\chi=0$, it
becomes the static wormhole metric~\cite{Morris:1988cz}, while for $%
\Phi(r)=b(r)=0$, the~metric (\ref{infWHmetric}) reduces to a flat de Sitter
metric. By~imposing $\Phi(r)\rightarrow 0,b/r\rightarrow 0$ as $%
r\rightarrow\infty$, the~spacetime becomes asymptotically de Sitter. As~in
the static Morris--Thorne solution, $\Phi(r)$ is everywhere finite so that
the only horizons present are cosmological. Note that the spacetime described by
Equation~(\ref{infWHmetric}), unlike the usual flat de Sitter spacetime, is
inhomogeneous due to the presence of the~wormhole.

The primary goal in the Roman analysis was to use inflation to enlarge an
initially small~\cite{Roman:1992xj}, possibly submicroscopic, wormhole. $%
\Phi(r)$ and $b(r)$ are chosen to give a reasonable wormhole at $t=0$, which
is assumed to be the onset of inflation. One can verify that the wormhole
expands in size by~considering the proper circumference $C$ of the wormhole
throat, $r=r_0,$ for $\theta=\pi/2$, at~any time $t=\mathrm{const}$
\begin{equation}
C=\int_0^{2\pi} e^{\chi t}\, r_0\,d\phi =e^{\chi t}\,(2\pi r_0)\,,
\end{equation}
which is simply $e^{\chi t}$ times the initial circumference. The~radial
proper length between two points $A$ and $B$, at~any $t=\mathrm{const}$, is
similarly given by Equation~(\ref{evolvproperdistance}), which reduces to
\begin{equation}
l(t)=\pm e^{\chi t} \int_{r_A}^{r_B} {\frac{dr}{{[1-b(r)/r]^{1/2}}}}\,,
\end{equation}
which is $e^{\chi t}$ times the initial radial proper separation. Thus, both
the size of the throat and the radial proper distance between the wormhole
mouths increase exponentially with time. It is fairly straightforward to
verify that the ``wormhole form'' of metric (\ref{infWHmetric}) is preserved
with time by~using the analysis following Equation~(\ref{barredslice}) and~by
replacing the conformal factor with the scale factor, $R(t)=e^{2\chi t}$.

To determine the stress--energy tensor necessary to sustain the wormhole
described by Equation~(\ref{infWHmetric}), it is useful to consider the set of
orthonormal basis vectors defined by
\begin{eqnarray}
\mathbf{e}_{\hat{t}}&=&e^{-\Phi}\,\mathbf{e}_{t} ,  \notag \\
\mathbf{e}_{\hat{r}}&=&e^{-\chi t}\,(1-b/r)^{1/2}\,\mathbf{e}_{r},  \notag \\
\mathbf{e}_{\hat{\theta}}&=&e^{-\chi t}\,r^{-1}\,\mathbf{e}_{\theta},  \notag
\\
\mathbf{e}_{\hat{\phi}}&=&e^{-\chi t}\,(r\,\sin\theta)^{-1}\,\mathbf{e}%
_{\phi}.
\end{eqnarray}
Thus, the~stress--energy tensor components, $T_{\hat{\mu}\hat{\nu}}$,
obtained from Equations~(\ref{dynTtt})--(\ref{dyn-p}) are \linebreak  given by

\begin{eqnarray}
T_{\hat t \hat t}&=&\rho(r,t) ={\frac{1}{{8\pi}}}\left(3{\chi}^2\,e^{-2\Phi}
+\,e^{-2\chi t}\,\,{\frac{b^{\prime }}{{r^2}}}\right) \,,  \label{infWHrho}
\\
T_{\hat r \hat r}&=&-\tau(r,t) ={\frac{1}{{8\pi}}}\left\{-3{\chi}%
^2\,e^{-2\Phi} -e^{-2\chi t}\,\left[{\frac{b}{{r^3}}}-{\frac{{2{\Phi}%
^{\prime }}}{r}}\, \left(1-{\frac{b}{r}}\right)\right]\right \} \,,
\label{infWHtau} \\
T_{\hat t \hat r}&=&-f(r,t) ={\frac{1}{{8\pi}}}\left[2e^{-\Phi-{\chi t}}\,
\left(1-{\frac{b}{r}}\right)^{1/2}\,{\chi}\,{\Phi}^{\prime }\right] \,,
\label{infWHflux} \\
T_{\hat \theta \hat \theta}&=&T_{\hat \phi \hat \phi} =p(r,t)  \notag \\
&=&{\frac{1}{{8\pi}}}\biggl\{-3{\chi}^2\,e^{-2\Phi}\,+ \,e^{-2\chi t}\,%
\biggl[{\frac{1}{2}}\left({\frac{b}{{r^3}}}- {\frac{{b^{\prime }}}{{r^2}}}%
\right) +{\frac{{{\Phi}^{\prime }}}{r}}\,\left(1-{\frac{b}{2r}}- {\frac{%
b^{\prime }}{2}}\right)  \notag \\
&&+\left(1-{\frac{b}{r}}\right) \left({\Phi}^{\prime \prime }+({\Phi}%
^{\prime 2 }\right) \biggr]\biggr \} \,,  \label{infWHp}
\end{eqnarray}
where the quantities $\rho,\,\tau\,,f$, and~$p$ are defined as before,and~are finite for all $t$ and $r$. Note from Equation~(\ref{infWHflux}) that
the flux vanishes at the wormhole throat. If~$\Phi(r)\rightarrow
0,\,b/r\rightarrow 0$ as $r\rightarrow\nobreak\infty$, the~stress--energy
tensor components asymptotically assume their de Sitter forms, i.e.,~$T_{\hat
t\hat t}=-T_{\hat r\hat r}=-T_{\hat \theta\hat \theta}=-T_{\hat \phi\hat
\phi}=3{\chi}^2$.

A particularly simple example of an inflating wormhole is obtained by
setting a null redshift function $\Phi(r)=0$ \cite{Roman:1992xj} so that
the stress--energy tensor components in an orthonormal frame reduce to
\begin{eqnarray}
T_{\hat t \hat t}&=&\rho(r,t) ={\frac{1}{{8\pi}}}\left[3{\chi}^2\,
+\,e^{-2\chi t}\,\,{\frac{b^{\prime }}{{r^2}}}\right] \\
T_{\hat r \hat r}&=&-\tau(r,t) ={\frac{1}{{8\pi}}}\left[-3{\chi}^2\,
-e^{-2\chi t}\,\left({\frac{b}{{r^3}}}\right)\right] \\
T_{\hat t \hat r}&=&-f(r,t)=0 \\
T_{\hat \theta \hat \theta}&=&T_{\hat \phi \hat \phi} =p(r,t) ={\frac{1}{{%
8\pi}}}\biggl[-3{\chi}^2\,+ \,{\frac{{e^{-2\chi t}}}{2}}\left({\frac{b}{{r^3}%
}}- {\frac{{b^{\prime }}}{{r^2}}}\right)\biggr].
\end{eqnarray}
Note that the stress--energy tensor components all approach their de Sitter
space values for a large value of $t$. When $\chi=0$, the solution reduces to that of a
static ``zero-tidal-force'' wormhole~\cite{Morris:1988cz}.

\subsection{The Properties of the Solutions and NEC~Violation}

Analyzing the stress--energy components of an evolving wormhole, in~particular those of an inflating wormhole, one verifies that a noticeable
difference between the properties of a generic redshift function with $%
\Phi(r)\neq 0$ and that of $\Phi=0$ wormholes is the presence of a flux
term, given by Equation~(\ref{infWHflux}). Following the Roman analysis \cite%
{Roman:1992xj}, to~clarify this difference, consider two coordinate systems
associated with the wormhole. The~first can be thought of as the rest frame
of the wormhole geometry, i.e.,~an observer at rest in this frame is at
constant $r,\,\theta,\,\phi$. The~second can be associated with the rest frame
of the wormhole material. One can define the rest frame of the wormhole
material as that in which an observer co-moving with the material sees
zero energy flux, and~from Equation~(\ref{infWHflux}), one verifies that for $%
\Phi(r)\neq 0$, the~wormhole material is not at rest in the $%
r,\,\theta,\,\phi$ coordinate system. For~the $\Phi(r)=0$ metrics, the~two
coordinate systems~coincide.

To analyze certain interesting properties of these wormholes, consider the
four-velocity of an observer who remains at rest with respect to the $%
r,\,\theta,\,\phi$ coordinate system, given by $U^{\mu}=dx^{\mu}/{d\tau}=
(e^{-\Phi(r)},0,0,0)$. The~observer's four-acceleration, $a^{\mu}=U^{%
\mu}{}_{;\,\nu}\,U^{\nu}$, provides the following relations:\vspace{-12pt}

\begin{eqnarray}
a^t&=&0  \notag \\
a^r&=&\Gamma^r_{tt}\,\left({\frac{dt}{{d\tau}}}\right)^2 =e^{-2\chi t}\,{\Phi%
}^{\prime }\,(1-b/r)\,.
\end{eqnarray}
From the geodesic equation, a~radially moving test particle which is
initially at rest has the equation of motion
\begin{equation}
{\frac{{d^{\,2}r}}{d{\tau}^2}}=-\Gamma^r_{tt}\, \left({\frac{dt}{{d\tau}}}%
\right)^2 =-a^r\,.  \label{inf:eqmotion}
\end{equation}
Therefore, $a^r$ is the radial component of proper acceleration that an
observer must maintain in order to remain at rest at constant $%
r,\,\theta,\,\phi$. Thus, from~Equation~(\ref{inf:eqmotion}), one deduces that for 
$\Phi(r)\neq 0$ wormholes, static or inflating, such observers do not move
along geodesics (except at the throat), contrary to $\Phi(r)=0$ wormholes.

The analysis of the ``attractive'' or``repulsive'' nature of the wormhole
geometry is analogous to that in the Morris--Thorne case. Additionally, the~sign of
the energy flux depends on the sign of ${\Phi}^{\prime }$, or~equivalently
on the sign of $a^r$. From~the flux term $f=-T_{\hat t \hat r}$, i.e.,~Equation~(%
\ref{infWHflux}), one verifies that if the wormhole is attractive, there is
a negative energy flow out of it; if it is repulsive, there is a negative
energy flow into~it.

To analyze the NEC, one may follow the Roman analysis and consider the
exoticity function. However, it is simpler to use the definition of the NEC
for a radial null vector. Consider $k^{\hat{\mu}}=(1,\pm 1,0,0)$, a~radial
outgoing (ingoing) null vector, so that from Equations~(\ref{infWHrho})--(\ref%
{infWHp}), one has
\begin{equation}
T_{\hat{\mu}\hat{\nu}}\,k^{\hat{\mu}}\,k^{\hat{\nu}}= {\frac{e^{-2\chi t}}{%
8\pi}} \biggl[\biggl({\frac{b^{\prime }}{{r^2}}}-{\frac{b}{{r^3}}}\biggr) -{%
\frac{2{\Phi}^{\prime }}{r}}\,\biggl(1-{\frac{b}{r}}\biggr)\biggr] \pm{\frac{%
e^{-\chi t}}{4\pi}} \biggl[\biggl(1-{\frac{b}{r}}\biggr)^{1/2}\, \chi\,{\Phi}%
^{\prime -\Phi}\biggr]\,.  \label{infWEC}
\end{equation}
For $\Phi(r)=0$, Equation~(\ref{infWEC}) reduces to
\begin{equation}
T_{\hat{\mu}\hat{\nu}}\,k^{\hat{\mu}}\,k^{\hat{\nu}} ={\frac{e^{-2\chi t}}{%
8\pi}} \biggl({\frac{b^{\prime }}{{r^2}}}-{\frac{b}{{r^3}}}\biggr)\,.
\end{equation}

From Equation~(\ref{infWEC}), one verifies that $(1-b/r)\rightarrow 0$ at the
throat, considering the finite character of $\Phi^{\prime }$, and~taking
into account the flaring-out condition~\cite{Morris:1988cz}, the~NEC is
violated at or near the throat $r=b=r_0$. If~$\rho$ is non-zero and finite
for all $t$, then the NEC at the throat decays exponentially with~time.

Roman~\cite{Roman:1992xj} went on to explore interesting properties of
inflating wormholes, in~particular by~analyzing the constraints placed on the
initial size of the wormhole if~the mouths were to remain in causal contact
throughout the inflationary period and the maintenance of the wormhole
during and after the decay of the false vacuum. It is also possible that the
wormhole will continue to be enlarged by the subsequent FRW phase of
expansion. One could perform a similar analysis to ours by replacing the
de Sitter scale factor in Equation~(\ref{infWHmetric}) with an FRW scale factor $%
a(t) $, which will be applied in the next~section.

\section{Evolving Cosmological Wormholes in Hybrid Metric--Palatini~Gravity}\label{sectionIV}

The most natural extension of general relativity is considering curvature
invariants in a gravitational Lagrangian.~
$f\left( R\right) $ gravity is one
of the most famous examples of these kinds of theories. Considering the metric
as a fundamental field and varying the action with respect to it, one
receives metric $f\left( R\right) $ gravity~\cite{Nojiri:2010wj,Sotiriou:2008rp,Capozziello:2011et,Bamba:2015uma,Nojiri:2017ncd}. This theory introduces
additional scalar degrees of freedom, which should have a low mass in order
to suitably explain the dynamics of the cosmos at a large scale.~
However, the~effects of this low-mass scalar field should be detectable at small scales,
such as in the Solar system, whereas there is no evidence for this~\cite{Capozziello:2007eu,Khoury:2003rn}. From~the viewpoint
of the Palatin formalism and by varying the $f\left( R\right) $ action with
respect to the metric and an independent connection~\cite{Olmo:2011uz}, there is no longer an
additional degree of freedom since the scalar field is an algebraic function
of the trace of the energy-momentum tensor. This also, in~itself, brings a
problem because it resorts to an infinite tidal force on the surface of
massive astrophysical-type objects~\cite{Olmo:2011uz}.

The problems pointed out provide enough motivation to introduce a
more compatible new framework. One idea is to introduce a hybrid metric--Palatini formalism~\cite{Harko:2011nh}.
On the one hand, this formalism introduces a long-range force into its scalar--tensor representation
and therefore automatically passes local measurements such as those at the level of the Solar system.
On the other hand, studies show that this hybrid theory explains cosmological
epochs properly~\cite{Lima:2014aza,Lima:2015nma,Harko:2018ayt}.
Thus, this hybrid metric--Palatini theory is a promising context for astrophysical~\cite{Capozziello:2012qt,Capozziello:2013uya,Avdeev:2020jqo} and
cosmological~\cite{Capozziello:2012ny,Boehmer:2013oxa,Lima:2014aza,Lima:2015nma,Fu:2016szo,Bronnikov:2020zob} studies and also for perusing compact objects, such as black holes~\cite{Bronnikov:2019ugl,Chen:2020evr,Rosa:2020uoi,Bronnikov:2020vgg} and
wormholes~\cite{Rosa:2018jwp,Capozziello:2012hr}.
\subsection{Action and Field Equations of Hybrid Metric--Palatini~Gravity}

The action of hybrid metric--Palatini gravity is given by%
\begin{equation}
S=\frac{1}{16\pi G}\int d^{4}x\sqrt{-g}\left[ R+f(\mathcal{R})\right]
+S_{m}\ ,  \label{eq:S_hybrid}
\end{equation}%
where $R$ is the metric Ricci scalar, and $S_{m}$ is the matter action. $%
\mathcal{R}\equiv g^{\mu \nu }\mathcal{R}_{\mu \nu }$ is the Palatini
curvature in which the Palatini Ricci tensor $\mathcal{R}_{\mu \nu }$ is
given by
\begin{equation}
\mathcal{R}_{\mu \nu }\equiv \hat{\Gamma}_{\mu \nu ,\alpha }^{\alpha }-\hat{%
\Gamma}_{\mu \alpha ,\nu }^{\alpha }+\hat{\Gamma}_{\alpha \lambda }^{\alpha }%
\hat{\Gamma}_{\mu \nu }^{\lambda }-\hat{\Gamma}_{\mu \lambda }^{\alpha }\hat{%
\Gamma}_{\alpha \nu }^{\lambda }\,,
\end{equation}%
where $\hat{\Gamma}_{\mu \nu }^{\alpha }$ is an independent connection. The~scalar--tensor representation of this gravity is~\cite{Harko:2011nh}%
\begin{equation}
S=\frac{1}{16\pi G} \int d^{4}x\sqrt{-g} \left[ (1+\phi )R+\frac{3}{2\phi }%
\partial _{\mu }\phi \partial ^{\mu }\phi -V(\phi )\right] +S_{m}.
\label{eq:S_scalar2}
\end{equation}%
From a computational point of view, the~above representation is clearly easier
to handle. The~field equations can be obtained by varying the action (\ref%
{eq:S_scalar2}) with respect to the metric and the scalar field, as%
\begin{gather}
G_{\mu \nu }=\kappa ^{2}\left( \frac{1}{1+\phi }T_{\mu \nu }+T_{\mu \nu
}^{(\phi )}\right) ,  \label{einstein_phi} \\ ~ \notag \\
\square \phi +\frac{1}{2\phi }\partial _{\mu }\phi \partial ^{\mu }\phi +%
\frac{\phi \lbrack 2V-(1+\phi )V_{\phi }]}{3}=\frac{\phi \kappa ^{2}}{3}T\ ,
\label{eq:evol-phi}
\end{gather}%
where 
 $T_{\mu \nu }$ and $T_{\mu \nu }^{(\phi )}$ are the energy-momentum
tensors of matter and the scalar field, respectively, and%
\begin{equation}
\label{tens_perfect}
T_{\mu \nu }^{(\phi )} =\frac{1}{\kappa ^{2}}\frac{1}{1+\phi }\Bigg[\nabla
_{\mu }\nabla _{\nu }\phi -\frac{3}{2\phi }\nabla _{\mu }\phi \nabla _{\nu
}\phi +\Bigg(\frac{3}{4\phi }\nabla _{\lambda }\phi \nabla
^{\lambda }\phi -\square \phi -\frac{1}{2}V\Bigg)g_{\mu \nu }\Bigg].
\end{equation}
\subsection{Traversable Cosmological Wormhole Solutions and Energy~Conditions}

In the framework of hybrid metric--Palatini gravity, four-dimensional traversable cosmological
wormholes have been studied~\cite{KordZangeneh:2020ixt}. Traversable
cosmological wormholes are shortcuts in spacetime that allow for human travel
and expand with the universe according to the same scale factor.
The exoticity of matter which supports wormhole spacetime has always provided grounds for doubt about
whether it is possible to have such structures in nature.
The results of~\cite{KordZangeneh:2020ixt} show
that in~the context of hybrid metric--Palatini gravity, there are four-dimensional wormhole structures supported completely
with normal matter. Such matter distributions satisfy both the null and weak energy conditions throughout space and at all times.
From one point of view, these results show that hybrid metric--Palatini gravity can be considered a rather more hopeful model and, from the other point of view, strengthens varying dark energy models because hybrid metric--Palatini gravity is one of them.
In what follows, we review this~kind of study.

The metric of a traversable cosmological wormhole can be written as%
\begin{equation}
ds^{2}=-dt^{2}+a^{2}\left( t\right) \left[ \frac{dr^{2}}{1-\frac{b\left( r\right)}{r}}+r^{2}\left( d\theta ^{2}+\sin ^{2}\theta d\varphi ^{2}\right) \right] ,
\label{met}
\end{equation}
where $b\left( r\right)$ is the shape function, and $a\left( t\right)$ is the scale
factor. The~wormhole structure has a throat $r_{0}$ which corresponds to a
minimum radial coordinate. In~order to have a traversable wormhole, the~shape function should satisfy the following conditions: \linebreak  (i
) $b(r_{0})=r_{0}$, (ii) $b(r)\leq r$ and (iii) $rb^{\prime }(r)-b\left(
r\right) <0$ (flaring-out condition) \cite{Morris:1988cz}. Here, we consider an anisotropic matter
energy-momentum tensor $T_{\nu }^{\mu }=\mathrm{diag}(-\rho ,-\tau ,p,p)$,
where $\rho $ is the energy density, $\tau $ is the radial tension, and $p$
is the tangential pressure. Then, we take into account the metric (\ref{met}) and
the gravitational field equations~(\ref{einstein_phi}), to~obtain the
following field equations:%
\begin{equation}
\rho (t,r)=\left( 3H^{2}+\frac{b^{\prime }}{r^{2}a^{2}}\right) \left( 1+\phi
\right) +3\dot{\phi}H+\frac{3\dot{\phi}^{2}}{4\phi }-\frac{1}{2}V(\phi ),~
\label{rhoeq}
\end{equation}%
\begin{equation}
\tau \left( t,r\right)  = \left( H^{2}+2\frac{\ddot{a}}{a}+\frac{b}{%
r^{3}a^{2}}\right) \left( 1+\phi \right) +2\dot{\phi}H-\frac{3\dot{\phi}^{2}}{4\phi }+\ddot{\phi}-\frac{1}{2}V(\phi
),
\end{equation}
and
\begin{equation}
p\left( t,r\right)  = \left( -H^{2}-2\frac{\ddot{a}}{a}-\frac{rb^{\prime }-b%
}{2r^{3}a^{2}}\right) \left( 1+\phi \right) -2\dot{\phi}H+\frac{3\dot{\phi}^{2}}{4\phi }-\ddot{\phi}+\frac{1}{2}V(\phi
),  \label{pteq}
\end{equation}
where $H=\dot{a}/a$ and $\phi =\phi \left( t\right) $. Note that the overdot
denotes the derivative with respect to the time coordinate, while the prime
denotes the derivative with respect to $r$. In~addition, we set $16\pi G=1$
for notational~simplicity.

In order to find the behavior of energy-momentum tensor components, we have to solve the equation of motion corresponding to a time-dependent scalar field (\ref{eq:evol-phi}) given by
\begin{equation}
\ddot{\phi}+3\dot{\phi}H-\frac{\dot{\phi}^{2}}{2\phi }-\frac{1}{3}\phi \left[
T+\left( 1+\phi \right) \frac{dV}{d\phi }-2V(\phi )\right] =0,
\label{phieom}
\end{equation}%
where $T$ is the trace of the energy-momentum tensor given by%

\vspace{-12pt}
\begin{adjustwidth}{-\extralength}{0cm}
\centering 
\begin{equation}
\label{trace}
T\left( t,r\right) = T_{\mu }^{\mu }=-\rho -\tau +2p
=-2\left( \frac{b^{\prime }}{a^{2}r^{2}}+3H^{2}+3\frac{\ddot{a}}{a}\right)
(1+\phi )-3\ddot{\phi}+\frac{3\dot{\phi}^{2}}{2\phi }-9\dot{\phi}H+2V(\phi ).
\end{equation}
\end{adjustwidth}
Since the scalar field $\phi$ depends only on time, the~differential Equation~(\ref{phieom}) should be independent of $r$.
To this end, the~trace $%
T\left( t,r\right) $ in this equation has to be $r$-independent. According
to (\ref{trace}), this leads to $b^{\prime }/r^{2}=CH_{0}^{2}$, where {$C$ is
an arbitrary dimensionless constant and} $H_{0}$ is the present value of the
Hubble parameter. {Consequently, the~shape function can be obtained as $%
b(r)=r_{0}+C{H_{0}^{2}}(r^{3}-r_{0}^{3})/3$, which clearly satisfies $%
b(r_{0})=r_{0}$. The~flaring-out condition at the throat yields $%
CH_{0}^{2}r_{0}^{2}<1$. In~what follows, we set }$C=0${\, which satisfies the
flaring-out condition and simplifies the solution.}

Now, we can solve the differential Equation~(\ref{phieom}) numerically for $%
\phi $. However, it is useful to re-express the equation in terms of
dimensionless functions of the scale factor $a$. Considering the following
definitions
\vspace{-12pt}
\begin{adjustwidth}{-\extralength}{0cm}
\centering 
\begin{equation}
\ddot{\phi} =\ddot{a}\phi ^{\prime }(a)+\dot{a}^{2}\phi ^{\prime \prime
}(a),\qquad \dot{\phi}=\dot{a}\phi ^{\prime }(a), \qquad \ddot{a} = H\dot{a}+\dot{H}a,\qquad \dot{H}=\dot{a}H^{\prime }(a),\qquad 
\dot{a}=Ha,
\end{equation}
\end{adjustwidth}
where the prime denotes the derivative with respect to the scale factor, $%
U=V/3H_{0}^{2}$, and~$E=H/H_{0}$, the~differential Equation~(\ref{phieom}) acquires
the final form below:%
\begin{equation}
\label{phidleom}
\phi ^{\prime \prime }(a)-\frac{\phi ^{\prime 2}}{2\phi (a)}+\frac{4(a\phi
^{\prime }(a)+\phi (a))}{a^{2}}+\frac{E^{\prime }(a)\phi ^{\prime }(a)}{E(a)}
+\frac{2\phi (a)E^{\prime }(a)}{aE(a)}-\frac{\phi (a)}{a^{2}E(a)^{2}}\frac{dU%
}{d\phi } =0.
\end{equation}
\textls[-15]{In the above differential equation, we have three unknown functions, namely $%
\phi (a)$, $E\left( a\right)$, and $U\left( \phi \right) $. Since we intend
to solve Equation~(\ref{phidleom}) for $\phi (a)$, we have to determine $E$ and $U
$. In~order to deduce $E$, we consider background quantities, $\rho
_{b}\left( t\right) =3H^{2}$ and $\tau _{b}\left( t\right) =H^{2}+2\ddot{a}/a
$ and a barotropic equation of the state $\tau _{b}=-\omega _{b}\rho _{b}$, which
leads to $E=a^{-3\left( \omega _{b}+1\right) /2}$ or, equivalently,
$a(t)\propto t^{2/3(1+\omega _{b})}$. Then, by~using dimensionless definitions, Equations~(\ref{rhoeq})--(\ref{pteq}) take the forms}
\begin{equation}
\frac{\rho }{3H_{0}^{2}}=\frac{1+\phi (a)}{a^{3\left( \omega _{b}+1\right) }}
+\frac{\phi ^{\prime }(a)}{a^{3\omega _{b}+2}}+\frac{\phi ^{\prime 2}(a)}{
4a^{3\omega _{b}+1}\phi (a)}-\frac{1}{2}U(\phi ),  \label{rho}
\end{equation}
\vspace{-30pt}
\begin{adjustwidth}{-\extralength}{0cm}
\centering 
\begin{equation}
\label{rhotau}
\frac{\rho -\tau }{3H_{0}^{2}}=\left( \frac{\omega _{b}+1}{a^{3\left(
\omega _{b}+1\right) }}-\frac{r_{0}}{3a^{2}H_{0}^{2}r^{3}}\right) \left(
1+\phi (a)\right)  +\frac{\left( \omega _{b}+1\right) \phi ^{\prime }(a)}{2a^{3\omega
_{b}+2}}+\frac{\phi ^{\prime 2}(a)}{2a^{3\omega _{b}+1}\phi (a)}-\frac{\phi
^{\prime \prime }(a)}{3a^{3\omega _{b}+1}},
\end{equation}
\end{adjustwidth}
and
\vspace{-18pt}
\begin{adjustwidth}{-\extralength}{0cm}
\centering 
\begin{equation}
\frac{\rho +p}{3H_{0}^{2}}=\left( \frac{\omega _{b}+1}{a^{3\left( \omega
_{b}+1\right) }}+\frac{r_{0}}{6a^{2}H_{0}^{2}r^{3}}\right) \left( 1+\phi
(a)\right) +\frac{\left( \omega _{b}+1\right) \phi ^{\prime }(a)}{2a^{3\omega
_{b}+2}}+\frac{\phi ^{\prime 2}(a)}{2a^{3\omega _{b}+1}\phi (a)}-\frac{\phi
^{\prime \prime }(a)}{3a^{3\omega _{b}+1}}.  \label{rhop}
\end{equation}
\end{adjustwidth}
Here, we consider $r_{0}=AH_{0}^{-1}$ for the wormhole throat so that the terms are dimensionless. We set the dimensionless constant $A$ to unity. Regarding the potential for the scalar field $U\left(\phi \right)$, we consider a quartic potential $\phi^4$. 

To construct wormhole structures, we follow a reverse procedure; namely,
we first build the geometry with the required conditions and then find the
energy-momentum profile. So, checking the energy conditions plays a
fundamental role. We expect that normal matter should satisfy the energy
conditions. The~weak energy condition (WEC) is defined as $T_{\mu \nu
}u^{\mu }u^{\nu }\geq 0$, where $u^{\mu }$ is a timelike vector. In~terms of the
components of the energy-momentum tensor, it is expressed as $\rho \geq 0$, $%
\rho -\tau \geq 0$, and $\rho +p\geq 0$. The null energy condition (NEC) is given
by $T_{\mu \nu }k^{\mu }k^{\nu }\geq 0$, where $k^{\mu }$ is \textit{any}
null vector. One could express this condition in terms of the energy-momentum
tensor components as $\rho -\tau \geq 0$ and $\rho +p\geq 0$, which are last
two inequalities in~the WEC.

In the following, we will review the satisfaction of the energy conditions for specific solutions with $\omega _{b}=1/3$ and $\omega _{b}=0$ where, respectively, $E=a^{-2}$ ($a(t) \propto t^{1/2}$) and $E~=~a^{-3/2}$ ($a(t)\propto t^{2/3}$).  These choices correspond tentatively to the radiation and matter~epochs.

\subsection{Specific Cases of $\protect\omega _{b}=1/3$ and $\protect\omega _{b}=0$}

\begin{figure}[H]
\begin{adjustwidth}{-\extralength}{0cm}
\centering 
\begin{tabular}{cc}
\includegraphics[width=.5\textwidth]{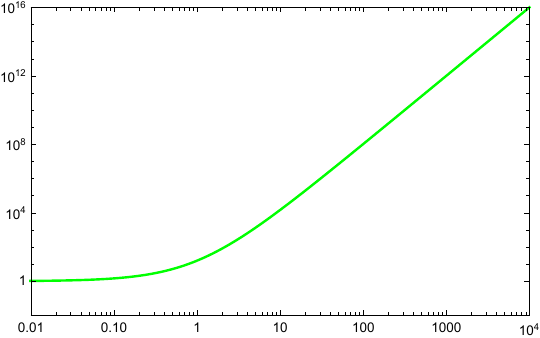}& \includegraphics[width=.5\textwidth]{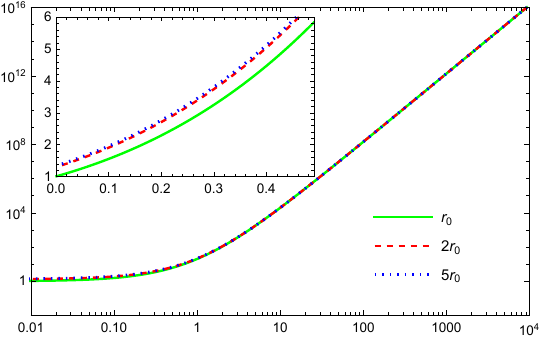} \\
({\bf a}) $\protect\rho/3H_0^2$ vs $z$&({\bf b}) $(\rho-\tau)/3H_0^2$ vs $z$\\
\multicolumn{2}{c}{\quad} \\
\multicolumn{2}{c}{\centering \includegraphics[width=.5\textwidth]{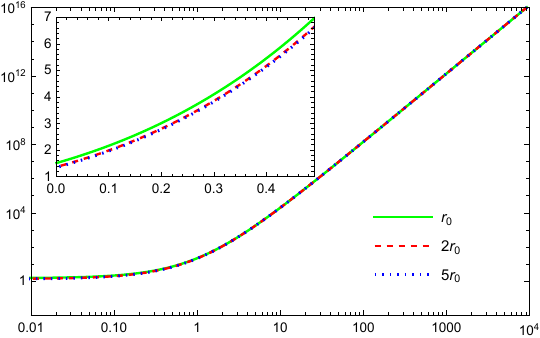}} \\
\multicolumn{2}{c}{\centering (\textbf{c}) $(\rho+p)/3H_0^2$ vs $z$} \\
\end{tabular}
\end{adjustwidth}

\caption{The behaviors of $\protect\rho $ (which is independent of $r$ according to Equation~(\ref{rho})), $\protect\rho -\protect\tau $, and $\protect\rho +p$, respectively, versus $z$ for different values of $r$ in
the specific case of $\protect\omega_b=1/3$ with $U(\protect\phi)=\protect%
\phi^{4}$. Note that both the horizontal and vertical axes are logarithmic.}
\label{fig3}
\end{figure}

In Figure~\ref{fig3}, the~behaviors of $\rho $, $\rho -\tau $, and $\rho +p$ with respect to the
redshift $z$ are displayed for $\omega _{b}=1/3$. These behaviors have been depicted for different values of $r$. Here, we have set the initial conditions for solving Equation~(\ref{phidleom}) numerically to
$\phi\left(\delta\right)=\phi^{\prime}\left(\delta\right)=10^{-4}$, where $\delta$ is very close to $a=0$. This~figure exhibits that these quantities are decreasing functions of time (or equivalently increasing functions of redshift), while they are always positive. Consequently, the NEC and the WEC remain satisfied at all~times.

\begin{figure}[H]
\begin{adjustwidth}{-\extralength}{0cm}
\centering 
\begin{tabular}{cc}
\includegraphics[width=.5\textwidth]{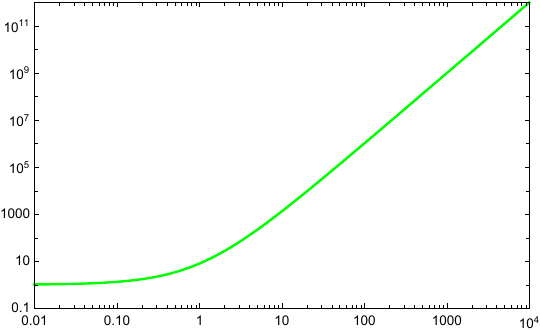}& \includegraphics[width=.5\textwidth]{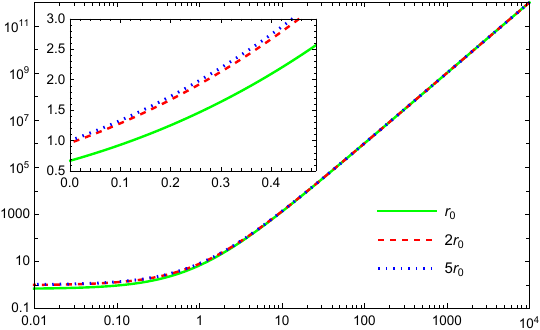} \\
({\bf a}) $\protect\rho/3H_0^2$ vs $z$&({\bf b}) $(\rho-\tau)/3H_0^2$ vs $z$\\
\multicolumn{2}{c}{\quad} \\
\multicolumn{2}{c}{\centering \includegraphics[width=.5\textwidth]{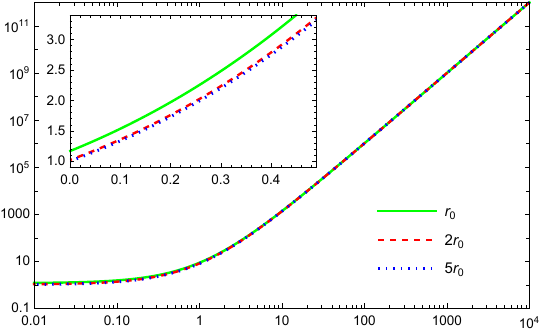}} \\
\multicolumn{2}{c}{\centering (\textbf{c}) $(\rho+p)/3H_0^2$ vs $z$} \\
\end{tabular}
\end{adjustwidth}



\caption{The 
 behaviors of $\protect\rho $ (which is independent of $r$ according to Equation~(\ref{rho})), $\protect\rho -\protect\tau $, and $\protect\rho +p$, respectively, versus $z$ for different values of $r$ in
the specific case of $\protect\omega_b=0$ with $U(\protect\phi)=\protect\phi%
^{4}$. Both horizontal and vertical axes are logarithmic.}
\label{fig4}
\end{figure}

Regarding the case $\omega _{b}=0$ and by setting the initial conditions to $\phi\left(\delta\right)=10^{-4}$ and $\phi^{\prime}\left(\delta\right)=-\frac{1}{2}$, the~energy conditions are depicted in Figure~\ref{fig4}. For~this case, the NEC and the WEC are satisfied at all times and for any radial coordinates $r$ as well. $\rho $, $\rho -\tau $, and $\rho +p$ also decrease (increase) with time (redshift).

\unskip

\section{Conclusions}\label{section:conclusion}

In this work, we have investigated the theoretical construction and physical viability of evolving wormhole solutions embedded within cosmological backgrounds, focusing on both general relativity and modified gravity frameworks, with~particular emphasis on hybrid metric--Palatini gravity. Motivated by the foundational insight that static traversable wormholes necessitate exotic matter that violates the classical energy conditions, we have explored time-dependent geometries as a means to minimize these violations and embed wormholes consistently within dynamically evolving and physically realistic cosmological models.
We reviewed and extended several classes of evolving wormhole solutions, including those conformally related to static metrics, wormholes sustained by minimally coupled scalar fields, and~configurations embedded into Friedmann–Lemaître–Robertson–Walker (FLRW) and de Sitter spacetimes. Our analysis confirmed that the dynamical character of the spacetime—particularly when influenced by cosmic expansion, scalar field evolution, or~modified gravity couplings—can play a crucial role in localizing or even eliminating violations of the null and weak energy conditions. In~particular, within~the framework of hybrid metric--Palatini gravity, we demonstrated that the additional scalar degrees of freedom introduced by the theory contribute effective geometric modifications to the stress--energy tensor, which can successfully mimic the role of exotic matter. This permits the construction of traversable wormholes that are supported entirely by matter fields obeying the NEC and the WEC across the full temporal evolution of the~universe.

These results represent a meaningful advancement toward identifying wormhole geometries that are not only mathematically consistent but also physically viable within the broader context of gravitational theory and cosmological modeling. By~embedding localized topological structures within an expanding universe and~incorporating generalized gravitational couplings, we have established new avenues for reconciling the theoretical existence of traversable wormholes with the constraints imposed by classical and semiclassical energy conditions. This line of inquiry significantly enhances our understanding of how non-trivial spacetime features can be naturally accommodated in cosmological scenarios.
Despite the progress achieved, several open questions and challenges remain. The~dynamical stability of these evolving wormhole solutions—especially under scalar, tensor, and~mixed-mode perturbations—remains an important topic of investigation, particularly in anisotropic or inhomogeneous cosmological settings. Furthermore, the~interplay between wormhole geometries and phenomena such as early-universe inflation, quantum gravitational effects, and~late-time cosmic acceleration deserves further attention. Understanding whether inflationary mechanisms can amplify microscopic wormholes to macroscopic sizes, or~whether quantum corrections might stabilize or destabilize such geometries, could provide key insights into the viability of wormholes as physical~objects.

Future work will aim to extend the classification of evolving wormhole solutions across a broader spectrum of alternative gravity theories, including $f(R)$, scalar–tensor, and~Gauss–Bonnet frameworks, as~well as higher-dimensional and compactified scenarios. The~inclusion of non-minimal couplings, anisotropic matter sources, and~rotating geometries will also be explored to capture the richness of the possible configurations better. Finally, investigations into the thermodynamic and quantum properties of these wormhole solutions—such as their entropy content, particle creation rates, and~semiclassical backreaction—may illuminate deeper aspects of gravitational dynamics and spacetime topology.
Ultimately, a~more complete understanding of how such non-trivial spacetime structures interact with the evolving fabric of the universe has the potential to shed light on profound and unresolved questions concerning the nature of gravity, the~topology of spacetime, and~the fundamental architecture of the~cosmos.

\vspace{6pt} 

\authorcontributions{All of 
 the authors have substantially contributed to the present~work. 
 Conceptualization, M.K.Z. and F.S.N.L.; methodology,  M.K.Z. and F.S.N.L.; software,  M.K.Z. and F.S.N.L.; validation,  M.K.Z. and F.S.N.L.; formal analysis,  M.K.Z. and F.S.N.L.; investigation,  M.K.Z. and F.S.N.L.; resources,  M.K.Z. and F.S.N.L.; writing---original draft preparation,  M.K.Z. and F.S.N.L.; writing---review and editing, M.K.Z. and F.S.N.L.; visualization,  M.K.Z. and F.S.N.L.; project administration, F.S.N.L.; funding acquisition, M.K.Z. and F.S.N.L. 
 All authors have read and agreed to the published version of the manuscript.}
\funding{This research was funded by Shahid Chamran University of Ahvaz under research grant no. SCU.SP1403.37271.
	This research was also funded by the Fundação para a Ciência e a Tecnologia (FCT) under research grants UIDB/04434/2020, UIDP/04434/2020, and PTDC/FIS-AST/0054/2021.
}

\dataavailability{
	Data are contained within the article.
} 
\acknowledgments{M.K.Z. thanks Shahid Chamran University of Ahvaz, Iran, for supporting 
 this work under research the grant no.~SCU.SP1403.37271.
F.S.N.L. acknowledges the support from the Funda\c{c}\~{a}o para a Ci\^{e}ncia e a Tecnologia (FCT) Scientific Employment Stimulus contract, with the reference CEECINST/00032/2018}



\conflictsofinterest{The authors declare no conflicts of~interest.} 


\begin{adjustwidth}{-\extralength}{0cm}

\reftitle{References}

\PublishersNote{}
\end{adjustwidth}
\end{document}